\PassOptionsToPackage{table,xcdraw,dvipsnames}{xcolor}

\documentclass[manuscript]{acmart}

\usepackage[table,xcdraw,dvipsnames]{xcolor}[usenames, dvipsnames]

\usepackage{tikz}
\usepackage{soul}
\usepackage{subfig}
\definecolor{lightgray}{gray}{0.96}

\newcommand*{\mybar}[1]{%
    \def\len{0.25}
    \pgfmathsetmacro\numnodes{\len*(#1/27.37)}
    \color{Orange}\rule{\numnodes\textwidth}{8pt}}
    
\newcommand*{\chart}[1]{%
  #1\% & \mybar{#1}%
}

\newcommand*{\chartthree}[1]{%
  #1\% & \mybarthree{#1}%
}

\newcommand*{\mybarthree}[1]{%
    \def\len{0.25}
    \pgfmathsetmacro\numnodes{\len*(#1/39)}
    \color{Orange}\rule{\numnodes\textwidth}{8pt}}

\usepackage{array}

\AtBeginDocument{%
  \providecommand\BibTeX{{%
    \normalfont B\kern-0.5em{\scshape i\kern-0.25em b}\kern-0.8em\TeX}}}

\setcopyright{acmcopyright}
\copyrightyear{2024}
\acmYear{2024}
\acmDOI{XXXXXXXXXXXXXXX}

\acmConference[TBD 'XX]{Not published in the ACM}{TBD, 2024}{TBD, TBD}
\acmBooktitle{TBD 'XX: Not published in the ACM,
  TBD, 2024, TBD, TBD}
\acmPrice{15.00}
\acmISBN{XXXXXXXXXXXXXXX}

\begin{document}


\title[Empirical Insights into Analytic Provenance Summarization]{Empirical Insights into Analytic Provenance Summarization: A Study on Segmenting Data Analysis Workflows}

\author{Shaghayegh Esmaeili}
\email{esmaeili@ufl.edu}
\orcid{0000-0002-1547-2700}
\affiliation{%
  \institution{University of Florida}
  \city{Gainesville}
  \state{Florida}
  \country{USA}
}

\author{Irelis D. Suarez}
\affiliation{%
  \institution{University of Florida}
  \city{Gainesville}
  \state{Florida}
  \country{USA}}
\email{irelis.suarez@ufl.edu}

\author{Ezekiel Ajayi}
\affiliation{%
  \institution{University of Maryland, Baltimore County}
  \city{Baltimore}
  \state{Maryland}
  \country{USA}}
\email{eajayi1@umbc.edu}

\author{Eric D. Ragan}
\affiliation{%
  \institution{University of Florida}
  \city{Gainesville}
  \state{Florida}
  \country{USA}}
\email{eragan@ufl.edu}

\renewcommand{\shortauthors}{Esmaeili et al.}

\begin{abstract}

The complexity of exploratory data analysis poses significant challenges for collaboration and effective communication of analytic workflows.
Automated methods can alleviate these challenges by summarizing workflows into more interpretable segments, but designing effective provenance-summarization algorithms depends on understanding the factors that guide how humans segment their analysis.
To address this, we conducted an empirical study that explores how users naturally present, communicate, and summarize visual data analysis activities.
Our qualitative analysis uncovers key patterns and high-level categories that inform users' decisions when segmenting analytic workflows, revealing the nuanced interplay between data-driven actions and strategic thinking.
These insights provide a robust empirical foundation for algorithm development and highlight critical factors that must be considered to enhance the design of visual analytics tools.
By grounding algorithmic decisions in human behavior, our findings offer valuable contributions to developing more intuitive and practical tools for automated summarization and clear presentation of analytic provenance.

\end{abstract}




\maketitle

\section{Introduction}\label{sec:introduction}

Many data analysis efforts take advantage of collaborative or distributed efforts with multiple people working together to understand data and make informed decisions.
In such cases, it is not only necessary to share data findings but also important to understand the analysis approach and steps taken to arrive at those findings~\cite{ragan2015characterizing,north2011analytic,xu2020survey}.
For example, in collaborative analysis, it is beneficial for analysts to understand each others' processes to reduce duplicated efforts and build off of one another's logic.
In larger analysis centers and team operations that adopt a division of labor, communication of logic and steps taken in the analysis is essential for reporting, verification, and quality control~\cite{madanagopal2019analytic}.

Communicating analysis processes to others is non-trivial for complex or open-ended data analysis scenarios, and the same is true for effectively understanding the workflow.
To assist these challenges, data exploration tools often include support for visualizing the provenance of analysis to help record the history of data operations, analyst actions, and findings (e.g.,~\cite{heer2008graphical,kurlander1988editable,ragan2015evaluating,dunne2012graphtrail}).
While detailed state history and event logs can provide a basis for extracting analytic provenance for visualization, data records and interaction logs are often too verbose or dense for easy human understanding~\cite{linder2016results}.
Effective communication needs summarization of key points of the analytic workflow.
To achieve this, a significant amount of manual human review, annotation, and summarization is typically required to support communication needs in practical work contexts~\cite{madanagopal2019analytic}.

Previous research has introduced several techniques and tools to facilitate generating visual presentations and to simplify communication of critical findings derived from analyses~\cite{kosara2013storytelling,mathisen2019insideinsights,gratzl2016visual}.
Recent advancements in large language models (LLMs), known for their proficiency in natural language summarization and contextual understanding, offer promising opportunities for enhancing automated provenance summarization.
Although methods exist for partially automating communication aids, generating human-meaningful and contextually appropriate summaries for effective communication and presentation of analytic provenance remains a significant challenge for automation algorithms.
Designing summarization techniques for people requires studying people to inform requirements.
Consequently, our research is motivated by the need for an empirical, human-centered foundation in designing automated summarization methods to enhance provenance visualization.

To this end, we conducted an empirical user study focused on the summarization and communication of analytical workflows in visual data analysis.
This study serves as a first step toward formulating human-based recommendations to configure algorithms for summarizing data provenance.
During the study, participants engaged in an intelligence analysis scenario using a visual data exploration tool. 
They first reviewed previous data analysis records and subsequently created summary note cards of the analytical workflows. 
We qualitatively analyzed the collected summary note cards to investigate how people summarize and break down an analysis workflow, as well as their preferred type of information and level of detail.
Our qualitative data analysis results suggest a set of keywords of significant importance and their patterns of change throughout a period of data analysis.
We also identified high-level categories of people’s underlying reasons and rationale when deciding how to separate summary stages of the provenance period.

\section{Related Work}

We expand the work on the background of analytic provenance, communication support for presentation and collaboration, and algorithmic techniques for automating summarization.

\subsection{Analytic Provenance}

A large body of prior work has contributed to research tools, techniques, and knowledge of provenance in the field of visualization.
Examples are highly varied in terms of research objectives and different uses of data history.
Many common examples of provenance-support features include basic history tracking, error recovery (i.e., undo and redo) to assist a user during data analysis, and recording data workflows to support replication of verification requirements (e.g.,~\cite{kurlander1988editable,freire2006managing,brodlie1993grasparc,derthick2001enhancing}).
Other applications such as event sequence analysis (see \cite{han2016flexible,guo2020survey}) may look at the provenance of interaction events with the goal of understanding patterns, trends, or notable behaviors (e.g.,~\cite{gotz2009characterizing, wongsuphasawat2012exploring}).
Along these lines, numerous studies have demonstrated the value of understanding an analysis through the inspection and visualization of interaction history~\cite{dou2009recovering,pena2019detecting,linder2016results}.
For example, Smuc et al.~\cite{smuc2009score} investigated how users develop insights in a hierarchical structure over time, emphasizing the evolving nature of insights in visual analytics workflows. 
Their work on participatory design highlights how tracking user insights can lead to a better understanding of user behavior.
Similarly, Stitz et al.~\cite{stitz2018knowledgepearls} introduced KnowledgePearls, which retrieves previous visualization states using provenance graphs to help users recall and revisit prior analysis sessions.
Though these works explore insight progression and interaction retrieval, our focus shifts toward user-driven workflow segmentation and its potential for automated summarization.
As another example with a different goal, other applications have employed techniques to summarize prior data coverage during analysis to help analysts monitor the breadth and completeness of their data exploration~\cite{feng2016hindsight,sarvghad2016visualizing}.

The emphasis of our research is analytic provenance, which primarily focuses on the human-oriented analysis and the history of user interactions and cognitive sensemaking during data analysis~\cite{north2011analytic,gotz2009characterizing,yan2021tessera}.
A review by Ragan et al.~\cite{ragan2015characterizing} organized different perspectives and focal areas of analytic provenance research to account for differences in (i) the types of history data considered and (ii) the various goals and purposes for its use in support of visual data analysis.
Another survey by ElTayeby and Dou~\cite{eltayeby2016survey} provided a review of the open-ended exploration of data with a focus on insights from user interaction log analysis in analytic visualization tools.
They identified five common steps used for interaction logs analysis in an analytic process: (i) interactions log recordings, (ii) encoding interactions and filtering, (iii) segmentation and chunking, (iv) automatic analysis and prediction, and (v) visualization techniques.
In our research, the most notable purposes are conducting a high-level meta-analysis of analytic processes and the supporting communication and presentation of past analyses.

Further advancing the organization of different concepts for provenance support, a recent survey by Xu et al. \cite{xu2020survey} adds consideration for (i) how the provenance data is encoded and (ii) how analytic techniques enable functionality to support those purposes.
The survey discusses the importance of the connection between the goals for provenance support and decisions for appropriate data representation and algorithmic techniques to support those goals.
Taking a different perspective, Battle and Heer~\cite{battle2019characterizing} provide a more holistic understanding of the exploratory visual analysis by reviewing relevant literature and highlighting key insights and themes.
They discuss how analysts' behavior can be predictable, and there are different structures in exploratory visual analysis regarding the depth of analysis. 
For example, analysts more often go down an exploration tree and iterate on existing questions rather than across by asking entirely new questions.

\subsection{Presentation, Narratives, and Collaboration}

For effective communication and interpretation of analytic provenance, summarization of analysis history is a known and well-established requirement for provenance research in visualization and interface design.
For instance, Kurlander and Feiner~\cite{kurlander1988editable} discussed using visual summary states in graphical histories to support branching workflows and undo/redo functionality.
In a later discussion of various techniques to support graphical provenance histories for visualization, Heer et al.~\cite{heer2008graphical} note the value of methods including history chunking, hierarchical organization, and interactive annotation.

The need for summarization and the challenges of data resolution of provenance information persist in operational data analysis environments.
Findings from a qualitative study by Madanagopal et al. \cite{madanagopal2019analytic} call attention to the different needs and priorities for the use of provenance information for different people depending on different job responsibilities and roles in professional data analysis environments.
Through interviews with experts covering a variety of professional areas (including intelligence analysis, cybersecurity, finance, scientific, and industry domains), their research identifies open challenges for provenance-support techniques to aid practical requirements such as quality control, analyst training, and organizational reporting.
To address operational challenges involving provenance data management and presentation capabilities, the study suggests a need for improved computational methods to assist with summarization and handling provenance data at variable levels of scope.

In a study of collaborative analysis and knowledge sharing among domain experts, Boukhelifa et al.~\cite{boukhelifa2019exploratory} present the benefits of the story format for sharing information from the data exploration processes among multiple experts.
Their study findings demonstrate the importance of process summarization into readily understandable narratives to facilitate common ground and communication.
In terms of asynchronous collaboration, Sarvghad et al.~\cite{sarvghad2015exploiting} also demonstrated that providing dimension coverage in tabular data, i.e., information about which dimensions were previously investigated by other people, would facilitate the analysis flow in collaboration, ``hand-off" of previous analysis, and reduce duplicate work.
Building on this, Block et al.~\cite{block2022influence} explored the impact of different visual provenance representations, such as data coverage and interaction history, on collaborative analysis. 
They found that data coverage representations allowed for more efficient exploration by summarizing what had been analyzed, while interaction history provided detailed past actions but often introduced additional cognitive load, slowing down the analysis process.

Varying summarization levels are also essential for many systems that incorporate provenance sharing to assist communication with the help of provenance information.
Collaborative data analysis involves multiple people working together to draw conclusions or insights from data.
As one example, Zhao et al.~\cite{zhao2017supporting} presented a design with \textit{Knowledge-Transfer Graphs} to support collaborator awareness of analysis progress by tracking the history of data relationships with a time-based visual graph.
With collaborative analysis tools supporting provenance, a summarized view can serve as an effective overview. However, collaborators might be interested in access to details on demand for selected aspects of the analysis history.
For example, in common node-link representations of graphical histories (e.g.,~\cite{freire2006managing,dunne2012graphtrail,javed2013explates}), nodes often serve as summary panels that visualize an intermediate output or step in the workflow.
With this design, the visual state of prominent nodes allows users to see the progression of analysis at a glance, while more details are typically available for closer inspection if desired.

In addition to collaborative analysis, other types of communication support target the need to bridge personnel serving different roles in the broader operational or organizational contexts \cite{madanagopal2019analytic}.
For instance, automated methods to assist reporting and presentation often aim to help a new audience (that is, people not actively or immediately engaged in conducting the analysis) understand the logic and rationale behind an analysis.
Related to goals for reporting and presentation are strategies for creating data narratives and adopting storytelling metaphors.
Segel and Heer~\cite{segel2010narrative} discussed the design space for data storytelling considering how graphical techniques and interactivity can support narrative structure and flow.
Gershon and Page \cite{gershon2001storytelling} argued for the value of adopting storytelling for visualization via techniques to enable fast and easy understanding of more complex information.
They discussed the benefit of reducing large amounts of information to smaller, more understandable points and the challenges of determining how to select concise and meaningful representations to create a narrative summary.

Others have explored the use of storytelling to support provenance presentations. 
For example, Wohlfart and Hauser introduced a method for volume visualization using storytelling for presentation~\cite{wohlfart2007story}.
In their proposed method, users can record interactions manually, modify recordings, and annotate them for storytelling purposes where the story can only be linear.
Expanding on these ideas, Kosara and Mackinlay~\cite{kosara2013storytelling} discuss methods for supporting the presentation of visualization logic and findings both through the main analysis workspace as well as with the aid of specialization presentation features such as slideshow refinement and annotation.

Numerous analysis tools support narrative authoring functionality, allowing manual crafting and annotation to help explain the findings or analysis process.
For example, research by Gratzl et al.~\cite{gratzl2016visual} provides an approach for unifying the data analysis environment with presentation tools by augmenting history tracking and story annotation.
In another work, Mathisen et al.~\cite{mathisen2019insideinsights} presented InsideInsights with an approach for hierarchical organization of narrative-oriented information to supplement the presentation of the analytic provenance for notable data findings.
With a focus on bridging the gap between visual analytics and storytelling, Chen et al.~\cite{chen2018supporting} present a method to assist story content creation as an intermediate between data analysis and story design.
Their proposed conceptual framework uses topic modeling to generate clusters focusing on different data dimensions, including time, space, linearly ordered values, and discrete categories.
These clusters can be selected as the starting point for the manual creation of story slides.

Taking a different model for representation, Park et al.~\cite{Park_2021} presented a data exploration system, StoryFacets, which uses a data flow model where each visual chart can be automatically reformatted into a presentation-oriented representation to aid communication.
Manual editing is essential for optimizing a narrative or presentation of an analysis workflow for effective communication, but algorithmic automation offers the potential to greatly reduce time and effort.

\subsection{Automating Provenance Summarization}

Effective presentation and communication designs use summarization to simplify long or complex data analysis activities.
To automate the generation of such presentations and communication aids from analytic provenance data, algorithmic techniques typically are needed for (i) segmentation of a sequence of data over time and (ii) content summarization for each segment. 
Provenance data is inherently time-based since it is based on the history or progression over time.
Therefore, it can benefit from the wide variety of existing techniques for simplifying and communicating temporal data.
For example, the adoption of existing methods for natural language processing, speech recognition, and topic modeling techniques are promising summarization techniques for temporally changing data, as demonstrated by Shi et al.~\cite{shi2018meetingvis} for summarization and visualization of the flow of group discussions during meetings.
Another example is EventFlow~\cite{monroe2013temporal}, which provides visual temporal summaries with the goal of simplifying event data from different record sets, as may be applicable when unifying different types of provenance data for an integrated presentation.
Alternatively, consider the ThemeRiver design~\cite{havre2000themeriver}, which uses topic modeling to augment a timeline to show the evolution of different data themes from textual data.
While some visualization designs provide a high-level overview of the flow of the analysis using metaphors such as storylines (e.g.,~\cite{chen2012visual}) or graph representations (e.g.,~\cite{freire2006managing}), we focus on designs that adopt a narrative approach for the presentation of provenance as a sequence of steps.

Examples by Linder et al.~\cite{linder2016results} and Mohseni et al.~\cite{mohseni2018provthreads} explored various techniques to segment a provenance timeline by analyzing changes in topic modeling as well as different indicators from user interaction.
They discuss challenges in selecting human-meaningful identifiers for change detection without the need for active human curation.
Separate efforts by Pena et al.~\cite{pena2019detecting} and Barczewski et al.~\cite{barczewski2020domain} demonstrated the techniques using Hidden Markov Models to detect changes in data analysis behaviors to help summarize and communicate analytic provenance.
In recent work, Yan et al.~\cite{yan2021tessera} proposed another strategy, Tessera, that focuses on sub-task and task segmentation for segmenting and modeling software event logs along with elements of the source data being analyzed.
This approach segments logs into blocks of goal-directed or task-oriented user behavior by estimating current intention and data coverage.

Also relevant to automating summarization by segmenting temporal data, a substantial body of work has developed myriad \textit{change point detection algorithms} for time-series data~\cite{aminikhanghahi2017survey}.
Many different change-detection approaches can use either supervised or unsupervised approaches to identify possible transition points in temporal data, allowing chunking provenance history into smaller, more human-manageable segments.
While supervised detection is an appealing prospect for automated summarization to circumvent the need for human parameterization or manual rule creation, it can be difficult to secure sufficient training data in many practical analysis scenarios, and we might expect communication design strategies to vary widely based on individual preferences.

LLMs, known for their ability to generate coherent summaries from complex textual data, could be integrated to improve the human understandability of provenance summaries.
LLMs can assist in identifying key insights and patterns from interaction logs~\cite{block2023summary}, providing a more nuanced understanding of analyst behavior.
However, despite the promise of LLMs, they also face limitations. 
These models can struggle with domain-specific jargon and may not always capture the deeper reasoning or context behind analyst decisions. 
Additionally, LLMs are prone to generating overly general or imprecise summaries without sufficient human oversight~\cite{burtsev2023working}. 
This highlights the need for continued research into combining algorithmic approaches with human expertise.

Thus, while a strong set of algorithmic tools is available to enable temporal simplification and summarization of provenance data, any implementation will likely still require human intervention and decision-making to effectively employ automation.
Similar challenges are shared across all algorithmic techniques regarding whether mathematically meaningful states translate to human-meaningful summarization states that match communication goals.
Our research aims to provide a human-centered basis for initial implementation or parameterization decisions.

\section{Research Goals}

The primary motivation of this research is to investigate user expectations for preferred types of information and the level of detail required to employ automated or semi-automated approaches for provenance reporting and presentation support.
While specific decisions and preferences for automated provenance summarization will likely depend on the specifics of any given analysis scenario, we expect many generalities to exist across different analysis contexts with respect to narrative strategies and the level of detail chosen for high-level presentation.
Thus, our research aims to provide an empirically grounded starting point for the implementation and parameterization of summarization techniques.
Furthermore, our study can provide a demonstration of qualitative analysis methodology as a template for others to contribute additional empirical data to guide the summarization and presentation design of analytic processes.

With a focus on supporting overview communications and higher-level narrative presentations, we designed a study to collect recommendations for data composition, level of detail, and temporal segmentation of analytic provenance workflows.
Therefore, in this paper, we aim to explore the following research goals:

\begin{itemize}
    \item What preferences for different types of provenance information do people adopt when summarizing data analysis processes?
    
    \item In segmenting the analysis process into smaller steps, what are the fundamental reasons and rationale used to decide on changes and to separate summary stages of the provenance timeline?
\end{itemize}

\section{Data Collection}

To address our research goals, we designed a study to understand how people would summarize and break down a data analysis workflow done by another person or another group of people.
The rationale for this decision is the need for communication and presentation of data analysis steps by people who are not necessarily analysts.
The analyst would probably have a better understanding of the steps they took to complete a data analysis workflow; however, the challenge is to understand how other people are summarizing the data analysis process with the goal of presenting or communicating steps and final results.
Therefore, we decided to use sample provenance data from a data repository of human analysis logs~\cite{mohseni2018analytic}.
Using sample provenance records from this repository, we conducted a new user study where our participants reviewed and summarized the provenance records. 

\begin{figure*}[t!]
    \centering
    \includegraphics[width=0.95\textwidth,height=0.16\textwidth]{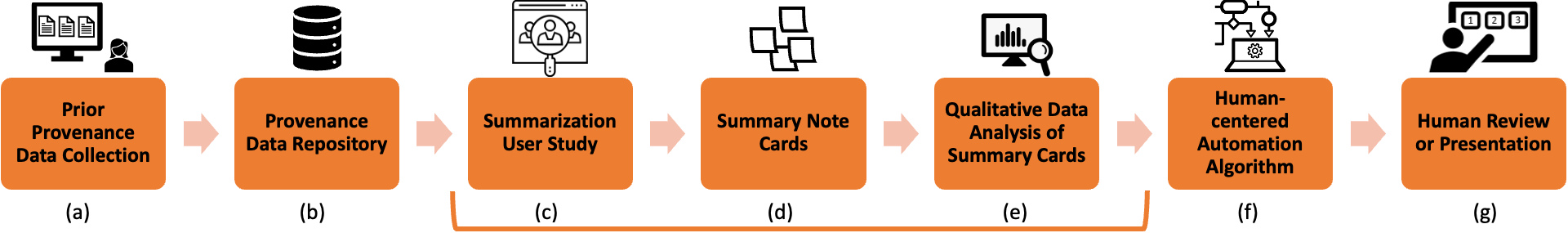}
    \caption{Overview of the main steps of this research are designated by the orange bracket within the context of the broader research goals: 
    (a) Prior user studies for sample provenance data.  
    (b) Sample provenance records were cleaned and aggregated to create a provenance data repository, including user interaction logs and think-aloud comments (section \ref{data-sample}).
    (c) A new user study of summarization of analytic workflows makes the core of the research in this paper.
    (d) We collected participants' summaries as note cards as the main data.
    (e) We qualitatively analyzed the collected data to extract patterns, themes, and features related to the summarization rationale (section \ref{analysis}).
    (f) Our research findings can inform the development and tuning of automated summarization algorithms.
    (g) This research will enable improved human-understandable summaries of analytic provenance for communication and presentation.
    The paper's focus is on steps (c), (d), and (e), and it provides fundamental results and themes for step (f).
    }
    \Description{A process flow chart consisting of the seven main steps describing the previous, current, and future steps of our research.
    There is a bracket in the figure indicating which steps are specifically covered in this paper, which are (c), (d), and (e).}
    \label{fig:overview}
\end{figure*}

Figure \ref{fig:overview} shows a high-level overview of the main steps of this research.

\subsection{Sample Provenance Data}
\label{data-sample}

\begin{figure}[t!]
 \centering
 \subfloat{
    \includegraphics[width=0.47\textwidth,height=0.29\textwidth]{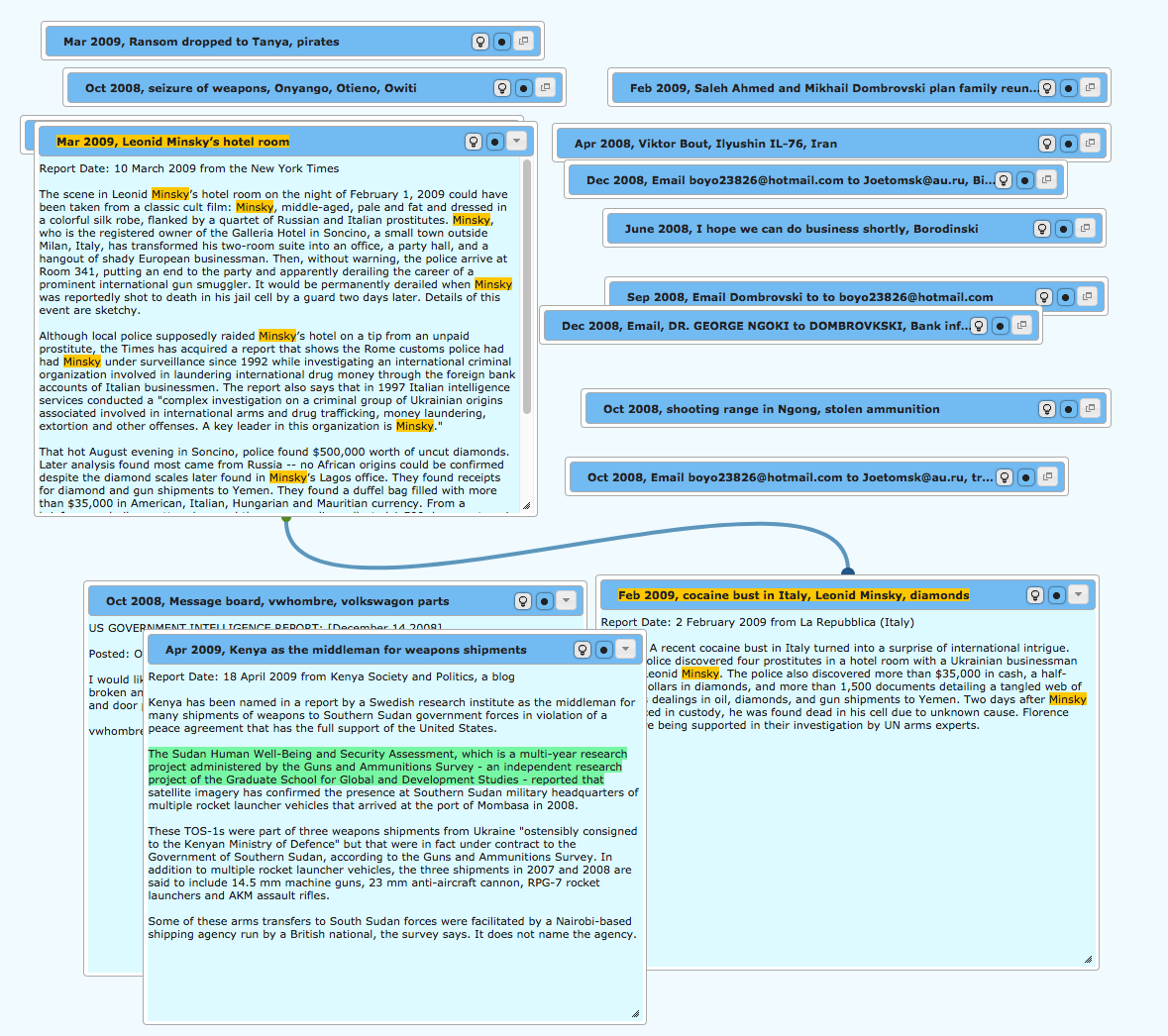}
 }
 \quad
 \subfloat{
    \includegraphics[width=0.47\textwidth,height=0.29\textwidth]{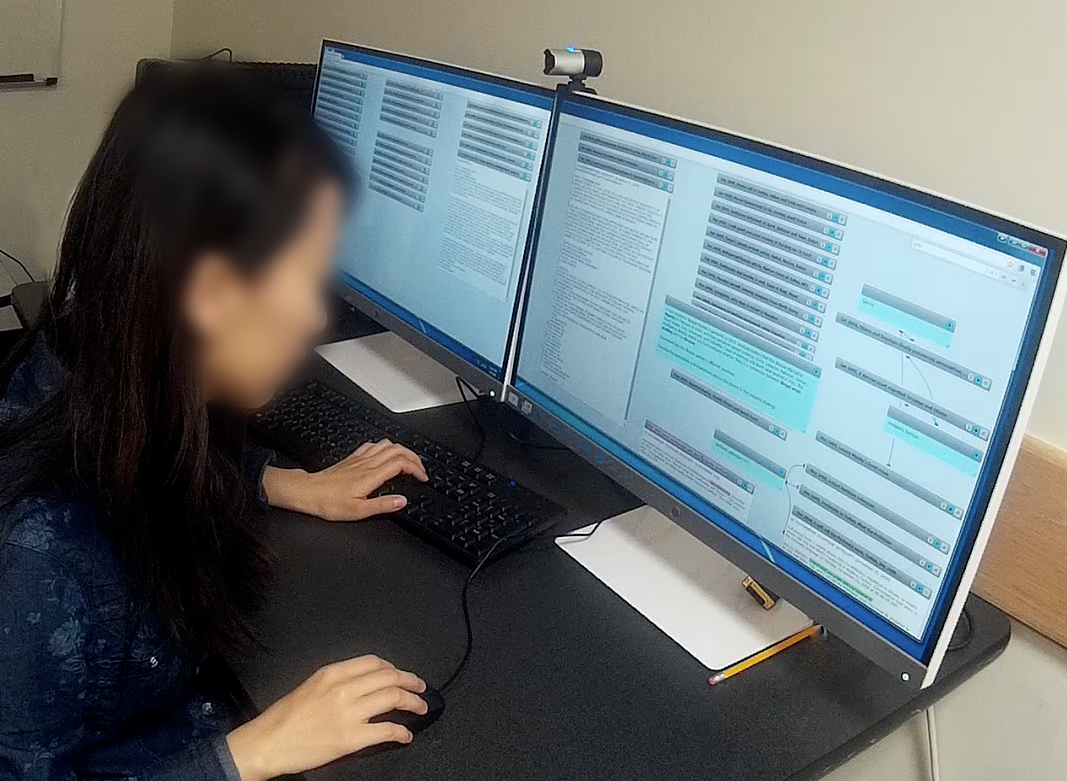}
 }

    \caption{
    The setting in which provenance data was collected for this paper's study (i.e., summarization user study).
    Left: Partial display of the visual provenance tool used for the data analysis scenario.
    Analysts are able to use it for reading documents, highlighting, searching, etc.
    The events and the analyst's actions would be recorded in the system background.
    Right: Prior study where an analyst would use the tool for a data analysis scenario. We used the recorded logs as sample provenance data in our summarization study.
    }
 \label{fig:vis-tool-current}

    \Description{The left-side figure displays the visual provenance tool consisting of some documents.
    Most documents are not open, i.e., only the title can be seen.
    However, there are four open documents, two of which have been connected with a link.
    There is also a sentence highlighted green, and the word Minsky'' has been highlighted yellow in various open documents and titles as a result of searching.
    The right-side figure shows a participant completing the prior study.
    There are two monitors in front of them displaying many documents, both open and closed, with some being connected and some text highlighted.
    The participant is using the mouse and the keyboard.}
\end{figure}

Data collection for the presented research had participants summarize provenance histories from prior analysis sessions.
We selected the sample provenance data from a provenance data repository~\cite{mohseni2018analytic} including users' interaction logs of a visual analytic system, their think-aloud comments, screen capture of the system during users' analysis process, and transcribed notes of users' activities.
This sample data was created through previous user studies where users used a basic visual document exploration tool (see Figure \ref{fig:vis-tool-current}) to analyze fictional intelligence analysis data.
The document visualization tool has basic text viewing functionality with the addition of text search, highlighting, annotation, and 2D document organization.

The provenance data repository contains analysis records based on sample analyses from multiple past IEEE VAST Challenges~\cite{cook2014vast}. 
In our user study (i.e., Provenance Summarization), we used past analysis records from the repository covering three textual intelligence analysis datasets from the 2010 mini-challenge 1, 2011 mini-challenge 3, and 2014 mini-challenge 1.
These different scenarios were selected for the purpose of replicability and generalizability of our current study.
In addition, for each of the scenarios, we selected two distinct provenance records (i.e., analysis records of two different analysts) to be used for summarization purposes in our study.
Therefore, the study consists of 6 provenance record sets (3 scenarios $\times$ 2 users' data analysis process).
Each participant completed the provenance summarization task with one of the 6 provenance samples.

\subsection{Study Task: Provenance Summarization}
\label{provenance-summarization}

The participant's task was to summarize and explain the key stages of an analytical process from the sample provenance records.
Participants were instructed to divide the entire duration of the data analysis workflow (embedded in the study platform) into smaller periods of time.
We made sure to communicate to the participant that it is ultimately their choice on how many stages they would have to summarize the data analysis process.
Moreover, we told them it was their choice to decide the duration of each stage summary as long as the total duration of stage summaries covered the whole duration of the data analysis process.

Our approach aligns with prior studies that elicited user preferences to guide system design. 
For instance, Bao et al.~\cite{bao2022recommendations}, interviewed public health experts to understand how they prioritize and approach visualization recommendations.
Their study revealed a preference for simple, clear visualizations tailored to specific domain needs, which parallels our efforts to understand how users segment and summarize analytic workflows. 
By incorporating user-driven feedback into our protocol, we ensure that our design is grounded in real-world preferences for workflow summarization, similar to how Bao et al. tailored recommendations based on expert input in their domain.

To facilitate the review of the sample provenance data, we developed a simple application that supported data browsing and ensured experimental consistency across the available tools.
The study tool consisted of a recorded video of a data analyst's computer screen as the ``sample analyst'' conducted the text analysis, as well as the sample data, which includes a temporal history of the analyst's verbal think-aloud comments and interactions events with the text visualization tool.

Since the videos were on average about 90 minutes, asking participants to fully watch, understand, and summarize the analysts' process within the duration of the study proved to be unfeasible during the pilot studies.
To alleviate this matter, we decided to modify the review tool in our study to link each timestamp and the corresponding description to the associated time in the video. Upon clicking on a transcript's timestamp, the video would adjust and play at the desired time.
Therefore, participants had the ability to not only see the whole video but also review it quicker by jumping around in the video playback based on their judgment of critical timestamps.

\subsection{Procedure} 
\label{procedure}

This study was approved by the university's Institutional Review Board (IRB).
We conducted an in-lab study, with one participant completing the study procedure at a time. 
Upon arrival, we obtained informed consent from participants before beginning the study.
Before the experimenter explained the study task and software, participants completed a brief background questionnaire about basic demographics, education, and data analysis experience.

To review the provenance records, participants used a computer and a monitor with a screen size of 27 inches and a resolution of 1920$\times$1080.
Participants interacted with the study tool through a computer mouse. The review software was developed as a web application run from a local server.

For participants to summarize the sample analysis into what they considered critical stages within the analysis period, the experimenter provided a large stack of 5$\times$8-inch paper index cards and pens. 
We made sure participants had more note cards available than would be reasonably needed or desired to ensure participants did not feel limited by supply during the summarization task. 

As part of the instructions, the experimenter provided an example card template containing a stage number, a start and end time for the proposed stage, and a placeholder for a summary description. 
However, the experimenter emphasized that the format in which the example card was presented was not a required structure for summarizing.
Participants were encouraged to utilize whatever approach they were comfortable with so long as their summaries could be easily segmented and contained an appropriate time frame, as well as their summary of the events transpiring during that time frame.

After completing the task, the experimenter would digitally photograph the index cards to capture their content and order.
The experimenter concluded the study session with a free-response interview. 
The purpose of the interview phase was to better understand the participant's rationale when summarizing, how they segmented the task, and whether or not they could condense their summaries when required.
The entire study procedure lasted approximately 90 minutes.
Breaks were provided as needed.

\subsection{Participants}
\label{participants}

A total of 25 students of graduate and undergraduate levels in different majors (such as computer science, digital arts, mathematics, and linguistics) participated in the study.
Two participants self-reported as female, and 23 reported as males. 
Ages ranged from 18 to 41 years, with a median age of 24.

When asked to rate their experience with data analysis, 76\% of the participants rated themselves 3 and 4 (\textit{Average} and \textit{Advanced}, respectively) on a 1--5 scale.
Their experiences were mostly qualitative and quantitative data analysis methods (such as data science-related or statistical-related). 
88\% of participants rated their experience with information visualization between 2 and 4 on a 5-point scale, with most prior experience involving basic data plotting and graph creation in Microsoft Excel.

None of the participants were familiar with the provenance data or the contexts used in the studies.

\section{Qualitative Data Analysis}
\label{analysis}

The nature of our research questions is to gather in-depth insights regarding human reasoning and thought processes in summarizing analytic provenance and further extract potential themes, concepts, and features for automatic summarization.
Accordingly, we decided to apply qualitative research methods to analyze our collected data.
Moreover, our collected data are in the form of descriptive observations and text-based summaries.
Therefore, qualitative data analysis is a well-suited approach for investigation purposes.

\begin{figure}[ht]
    \centering 
    \includegraphics[width=0.5\textwidth,height=0.45\columnwidth]{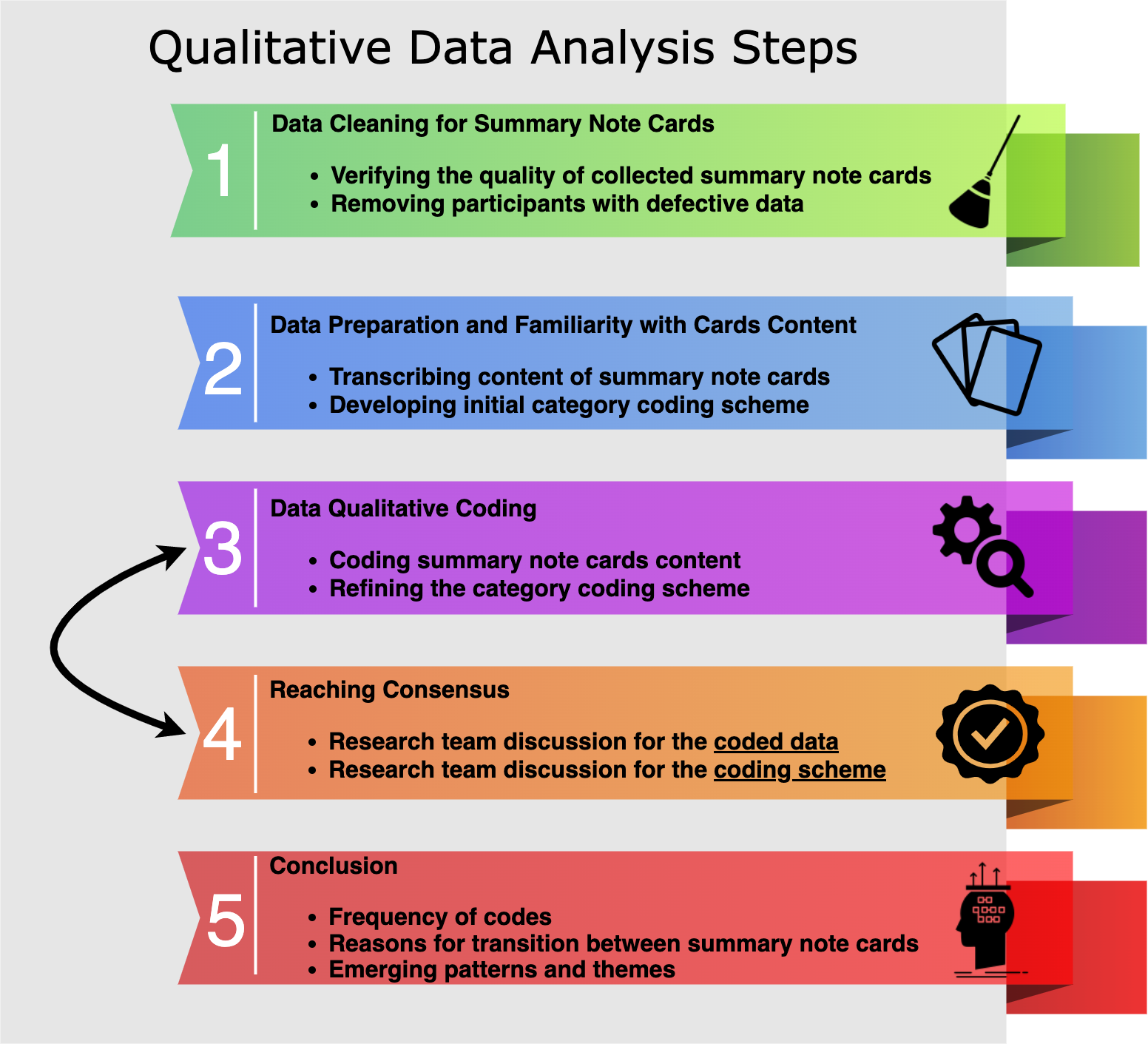}
    \caption{
    Overview of the main steps of our qualitative data analysis.
    The arrow between steps 3 and 4 indicates the iterative cycle between coding data and discussing them to reach an agreement. 
    This cycle ends when no further change is needed for either the coding scheme or coded data.
    }
    \label{fig:analysis-steps}
    \Description{
    This figure shows the steps we took for the qualitative data analysis.
    1. Data cleaning for summary note cards, including: 
    (i) verifying the quality of collected summary note cards. 
    (ii) removing participants with defective data.
    2. Data preparation and familiarity with cards content, including:
    (i) transcribing content of summary note cards.
    (ii) developing an initial category coding scheme.
    3. Data qualitative coding, including:
    (i) coding summary note cards content.
    (ii)refining the category coding scheme.
    4. Reaching consensus, including:
    (i) research team discussion for the \underline{coded data}.
    (ii) research team discussion for the \underline{coding scheme}.
    5. Conclusion, including:
    (i) frequency of codes.
    (ii) reasons for transition between summary note cards.
    (iii) emerging patterns and themes.
    As mentioned in the caption, there is an arrow between steps 3 and 4 indicating iterative cycles.
    }
\end{figure}

This section presents the steps we took for our qualitative data analysis.
Figure~\ref{fig:analysis-steps} shows a summary of the steps.

\subsection{Data Quality Verification}

We initiated the data analysis by assessing the quality of the gathered note cards containing each stage summary. 
We used specific criteria to evaluate the quality and relevance of participants' data for further analysis:
(1) Summaries needed to cover the entire video duration,
(2) Summary note cards needed to include clear timestamps for the beginning and end of each analysis period, and
(3) Summaries had to demonstrate that the participants comprehended the task and followed the study instructions.
Initially, we enrolled 25 participants; however, data from 7 participants were excluded based on these pre-defined quality criteria, resulting in a final sample size of 18 participants. 
This decision was made to ensure that only high-quality data were used for analysis, as these participants did not meet the requirements for task compliance, such as having inaccurate or missing timestamps or failing to cover the entire duration of the video.
The final selection also ensured a balanced number of participants in each condition.

\subsection{Qualitative Coding Scheme}

To analyze the content of the descriptive summary note cards, we first needed to develop a qualitative coding scheme that would allow us to extract important and relevant information to summarize data analysis workflows.
We started our coding scheme with initial categories of \textit{data}, \textit{interaction}, \textit{insight}, and \textit{rationale} (based on~\cite{ragan2015characterizing}).
We then finalized our coding scheme by going through our data iteratively and refining the categories.
The description of our code set and the categories is shown in Table \ref{tab:code-set} along with examples and frequency of observed keywords.

\begin{table*}[t!]
    \caption{Keyword categories and their Ratio. \textit{Keywords} are any word, term, phrase, or statement of great significance in note card content.
    \newline The total number of keywords across all participants is 1454.
    }
    \label{tab:coding-scheme}
    \centering\label{tab:code-set}
    \resizebox{\textwidth}{!}{
    \small
    \rowcolors{2}{}{lightgray}
    \begin{tabular}{m{0.08\textwidth}m{0.52\textwidth}m{0.15\textwidth}m{0.09\textwidth}m{0.04\textwidth}m{0.25\textwidth}} \hline & \\[-1.5ex]
        Keyword Category &
        Description &
        Example &
        Temporal Pattern &
        Ratio & \\ [1ex] \hline & \\[-1.5ex]
        
        Direct Data
        & Any word directly in the content of the source data sets. & \textit{Kenya, arm-trading}
        & rise fall
        & \chart{27.37}  \\[1ex] & \\[-1.5ex]
        
        Indirect Data
        & A reference to or an indirect description of the content of the source data sets.
        & \textit{this doc, the name}
        & fall
        & \chart{22.08} \\[1ex] & \\[-1.5ex]
        
        Action
        & Indicating any process of doing something or using tool facilities by the analyst.
        & \textit{write notes, search, organize}
        & fall
        & \chart{16.51}\\[1ex] & \\[-1.5ex]
        
        Strategy
        & Related to the higher-level behavior or action of the analyst as the strategy to solve the problem.
        & \textit{organizing based on events related to Kenya, search until another relevant article is found}
        & rise fall
        & \chart{11.83}\\ [1ex] & \\[-1.5ex]
        
        Other
        & Keywords that do not belong to any of the previous categories.
        & {[}decided based on the context{]}
        & variable
        & \chart{8.18}\\ [1ex] & \\[-1.5ex]
        
        Finding
        & Showing a significant and direct result of an action.
        & \textit{found important clue, found a piece of puzzle}
        & rise
        & \chart{5.85}\\ [1ex] & \\[-1.5ex]
        
        Insight
        & Showing a deeper and higher level understanding of the findings, i.e., a major clever result.
        & \textit{Nicolai and Dubai are the link}
        & rise
        & \chart{5.85}\\ [1ex] & \\[-1.5ex]
        
        Reasoning
        & Related to the purpose of an action and what it aims to achieve.
        & \textit{to establish correlations, to search for next steps}
        & rise fall
        & \chart{2.75}\\ [1ex] & \\[-1.5ex]
        
        Meta Comment
        & A clever or higher-level insight from the participant about the analyst's general behavior or data analysis procedure. 
        & \textit{comment: Even though it seems like a good idea, focusing only on ... may lead to wrong} direction.
        & rise fall
        & \chart{2.75} \\ [1ex] \hline
    \end{tabular}
    }
\end{table*}

A \textit{card} keyword is defined as \textbf{any} word, term, phrase, or statement in the note card, which is of great significance regarding our research goals. 
This category is the universal set of keywords, and any other types of keywords, such as \textit{data}, \textit{strategy}, etc., are a subset of \textit{card} keywords. 
In the initial version of our code set, we only considered \textit{direct data}, \textit{indirect data}, \textit{action}, \textit{findings}, \textit{reasoning}, and \textit{other} categories. 
However, as we continued going deeper into the data, more essential and higher-level categories emerged. 
Therefore, after a discussion between the research team members, we added \textit{strategy} and \textit{insight} keywords to the code set. 
Insight keywords would indicate a greater capacity to gain an accurate and deep intuitive understanding of the data analysis. 
Similarly, \textit{strategy} keywords refer to higher-order thinking, which leads to more intelligent behavior in solving a problem.

We also observed that some participants would write down their own thought processes and insights in the note cards, while there is no indication of the analyst's action or behavior in either the card content, video, or accompanying transcript. 
Thus, we added another category, \textit{meta comment}, to code such cases throughout our coding process.
For example, one participant wrote down the following at the end of their first note card: ``Comment/ even though it seems like a good idea, focusing only on the articles after Feb 2009 may lead to wrong directions,'' which indicates their critique of the analyst's approach.

Finally, we set some rules and assumptions in the coding scheme to ensure that all researchers are coding the data in a similar manner. 
For example, different categories in our code set (except \textit{card} keyword) are not mutually exclusive. 
In other words, one keyword can be coded in more than one category; for example, \textit{making a connection} can be interpreted and referred to both using the tool's ability to draw connections between articles, as well as a strategy to solve a problem by forming relations. 
The former would be coded as \textit{strategy}, and the latter as \textit{action}.

\subsection{Content Analysis}

Our data analysis goal was to identify how people summarized and segmented the video to present and communicate the history of an analysis process. 
Therefore, we conducted inductive qualitative data analysis to derive empirical information on how people think about and generate summaries of analysis periods. 
In this subsection, we describe the steps taken in our core qualitative analysis of participants summaries.

After the quality verification step, one researcher transcribed and digitized all summary cards (175 cards across 18 participants), which we considered our base for the next data analysis steps. 
For each card, we recorded:
\begin{itemize}
    \item Card number indicating positional ordering in the summary sequence
    \item Assigned provenance data set for the summary
    \item Timestamps for the beginning and end of covered card duration
    \item All summary text on the card
\end{itemize}
    
Two researchers separately went through the cards one by one to get more familiar with the data content and come up with keyword categories. 
Then, they met and shared their ideas about emergent important keywords, their categories, and themes, which led to the development of the initial code set.
    
In the first phase of coding, two researchers independently coded each card based on the first code set. 
Upon completing the coding process, they met and discussed their codes and categories to reach an agreement.
    
After reaching a consensus, the research team further discussed codes for particular cases, code set categories, and their definitions. 
The discussion led to refining the code set and adding extra codes based on the emerging types of keywords.
For example, initially, we only had \textit{data} category and would categorize any other terms such as ``this name'' and ``all people'' as \textit{other} category.
However, we observed that, though these terms are not directly in the data set, they indirectly refer to the content.
Therefore, we added \textit{indirect data} as a new category to differentiate these types of keywords from \textit{other} category and changed \textit{data} category to \textit{direct data}.
Table \ref{tab:code-set} shows the finalized code set.
    
Once we finalized the coding scheme, the second phase of coding started. 
The lead researcher iteratively coded all cards based on the refined code set. 
The final coding decision of note cards was based on the content of the cards for the most part. 
However, we also considered the events happening on the card's duration in both the video and transcript to have a better interpretation of the participant's summary.
    
Finally, the research team examined and discussed the frequencies of codes, the duration of cards, and reasons for changing cards to identify key bases for the strategies of the segmentation process and to understand the emerging patterns in codes.

\subsection{Analysis of Stage Transitions}

Automated summarization techniques need to not only account for the data content in summaries but also employ meaningful criteria for segmenting longer periods of analysis time into shorter, more easily understandable stages.
Therefore, to better inform ``breaking points'' to summarize stages of analysis, we analyzed the key reasons for change between any two consecutive summary cards. 
We defined \textit{key reasons for change} as what changed in the analyst's data analysis process between consecutive cards and why the participant switched to the next card in preparing the summary. 

Given this definition, two researchers independently reviewed the summary cards. 
They wrote down the reasons for change between each note card based on what they observed in the card content, the timestamp of change, and any important hints in the participant's interview responses. 
Afterward, the researchers met and discussed the reasons one by one and addressed any agreements and disagreements. 
Finally, the base data for the causes of change included 139 distinct reasons across all participants and their note cards. 

Next, we synthesized and organized the large number of reasons to reach higher level themes of key reasons for change between summary stages by using the affinity diagramming technique (also known as the KJ method~\cite{scupin1997kj}).
This common technique is useful for externalizing, making sense of, and giving meaning to large amounts of unstructured qualitative data~\cite{hartson2012}.
Once a consensus for key reasons for change was made, we labeled each reason and inductively grouped them to capture emergent features. 
One reason could belong to multiple themes if the reason were related to various theme ideas. 
The research team iteratively discussed and grouped the themes and sub-themes to reach an agreement on the themes, which capture the categories of different reasons, leading to summarizing the data analysis process into multiple stages.

\section{Results}
\label{sec:results}

In this section, we present the results and insights drawn from our qualitative data analysis. 
This section covers descriptive findings of workflow summaries and key reasons for change between summary note cards.

\subsection{Descriptive Findings of Summary Segmentation}

Automating the summarization of temporal data into smaller periods requires knowledge of the preferred number of stages and the duration covered by each.
Analyzing the summary cards from the study provides a foundational understanding of segmentation choices for summarizing the analysis records.

\begin{figure}[h!]
    \centering 
    \includegraphics[width=0.75\columnwidth]{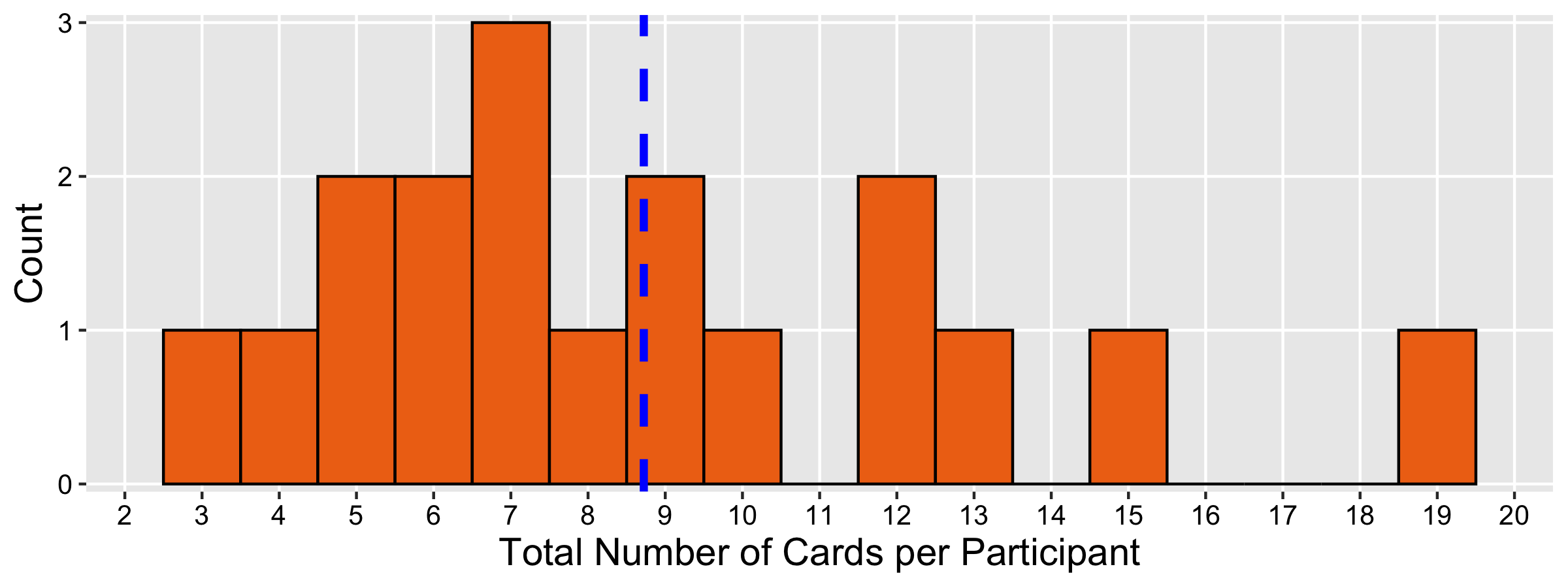}
    \caption{This histogram shows the total number of cards (per participant). 
    The blue dashed line shows the mean (8.7 cards).
    Across all participants, there were 157 summary cards.
    }
    \label{fig:numCards-participant}
    \Description{The total number of note cards per participant has been shown in a histogram format.
    The maximum count is 3 participants, who have seven cards in total.
    }
\end{figure}

Figure \ref{fig:numCards-participant} shows a histogram of the number of stages (i.e., indicated by distinct summary cards) that participants broke the analysis process into. 
The mean number of cards per participant was 8.7, and the median was 7.5. 
We observed that segmenting the data analysis process into seven stages is the most common across participants. Moreover, 77.78\% of participants used 4--12 cards to summarize the analysis.

Summarization approaches for temporal data can also be tuned based on segmentation preferences for the time covered in different segments.
Therefore, we studied the duration of analysis time summarized in each card from the study.
While the total time of the given analysis recordings was approximately 90 minutes, different approaches are possible for the duration and variance of sub-period times. 
The duration of summary cards from the study ranged from 0.06 to 48.80 minutes.
The shortest times were contributed by participants who had summarized the data analysis flow with a high granularity approach; these were the two participants with the highest numbers of total cards (15 and 19; see Figure~\ref{fig:numCards-participant}).
On the higher end, the duration times were skewed by a maximum duration belonging to a single note card covering a broad ``discovery/exploratory phase'' by the participant.
To account for the high variability in the level of detail and temporal coverage, we exclude outliers of durations from further analysis using the 1.5$\times$IQR rule (data points more than 1.5 times the interquartile range beyond the first or third quartiles).
Out of the 175 total cards, six were excluded as outliers. 
Figure \ref{fig:numCards-participant} (down) shows the duration distribution across all cards after removing outliers.
The interquartile range (i.e., the middle 50\% of the data) was between 1.85 minutes and 12.95 minutes.
While the variance was high, the distribution was clearly skewed.
The histogram shows a drastic decrease in the count of cards around the 3-minute mark; 41.06\% of all card durations were under 3 minutes.
Looking more into the variability of times, we found that the majority of summary cards with shorter duration (i.e., less than 3 minutes) indicate either the initial stages (such as the beginning of exploration and familiarity with documents) or the final stages (such as conclusions or wrapping up).
These results show that when people summarize the analysis flow, summary stages are not necessarily equal regarding the coverage of the analysis duration.
Another interesting observation is that the initial and final stages of data analysis have shorter durations.
In contrast, the middle stages, which are mainly dedicated to deeper analysis stages, have longer durations.

\begin{figure}[ht!]
    \centering 
    \includegraphics[width=0.75\columnwidth]{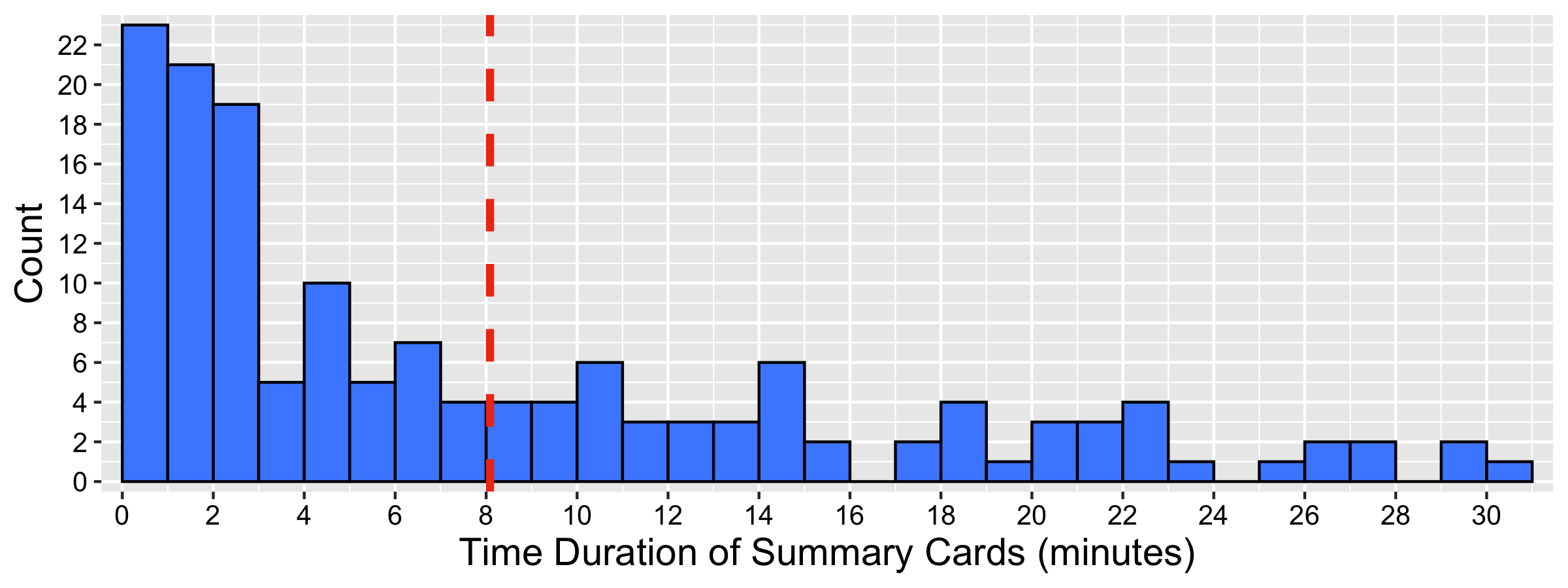}
    \caption{This histogram shows the duration of cards across all participants.
    The red dashed line shows the mean (8.08 minutes). 
    The min and max are 0.06 and 30.28 minutes.
    }
    \label{fig:duration}
    \Description{The time duration of note cards across all participants has been shown in a histogram format.
    The histogram has a decreasing trend, showing a drastic fall near the 3-minute mark.
    }
\end{figure}

\subsection{Summary Information by Keyword Categories}

Considering the total of 157 summary cards from all participants, the mean number of words used per card was approximately 23, and the median was 19.
We further analyzed the summary content to derive a more meaningful understanding of the types of information included in the summaries.
Content analysis of summary cards followed the qualitative coding scheme described in Section~\ref{analysis}, which led to the keyword categories summarized in Table~\ref{tab:coding-scheme}. 
Keywords made up approximately 40\% of the total written content, and participants' summaries averaged 9 keywords per summary card. 
We removed outliers in the count of keywords based on the 1.5$\times$IQR rule.

While identifying the types of information important for summarization is an essential first step, effective summarization methods will also need to give attention to the composition of the different types of information included in the summaries.
To this end, Table~\ref{tab:coding-scheme} (right) shows the ratio of the number of keywords in each category compared to the total number of keywords (note that keyword classifications are not mutually exclusive, thus allowing totals across all categories to exceed 100\%). 
The \textit{direct data} category has the highest number of keywords, which is 27.40\% of the total number of keywords ($398/1454$). 
\textit{Reasoning} and \textit{meta comment} categories have the lowest number of keywords which is 2.75\% of the total number of keywords ($40/1454$).

\subsection{Patterns in Keyword Categories over Time}    

In addition to analyzing the total composition of summary information by keyword categories, we examined the composition over time.
The keyword pattern results show that the majority of participants ($>$50\%) had \textit{direct data}, \textit{indirect data}, \textit{action}, and \textit{strategy} keywords in almost all of their cards over the analysis time.
This is consistent with the results related to the ratio of categories in Table~\ref{tab:coding-scheme} regarding the importance and high proportion of these keywords.

Moreover, we investigated temporal patterns based on the number of keywords for the different categories.
To do this, we calculated the ratio of each keyword category per card for each individual participant.  
Then, we investigated the changes in each keyword category over the course of all summary cards per participant to examine the relation between increasing or decreasing patterns with each category of keywords.

\begin{figure}[ht]
    \centering 
    \includegraphics[width=\columnwidth]{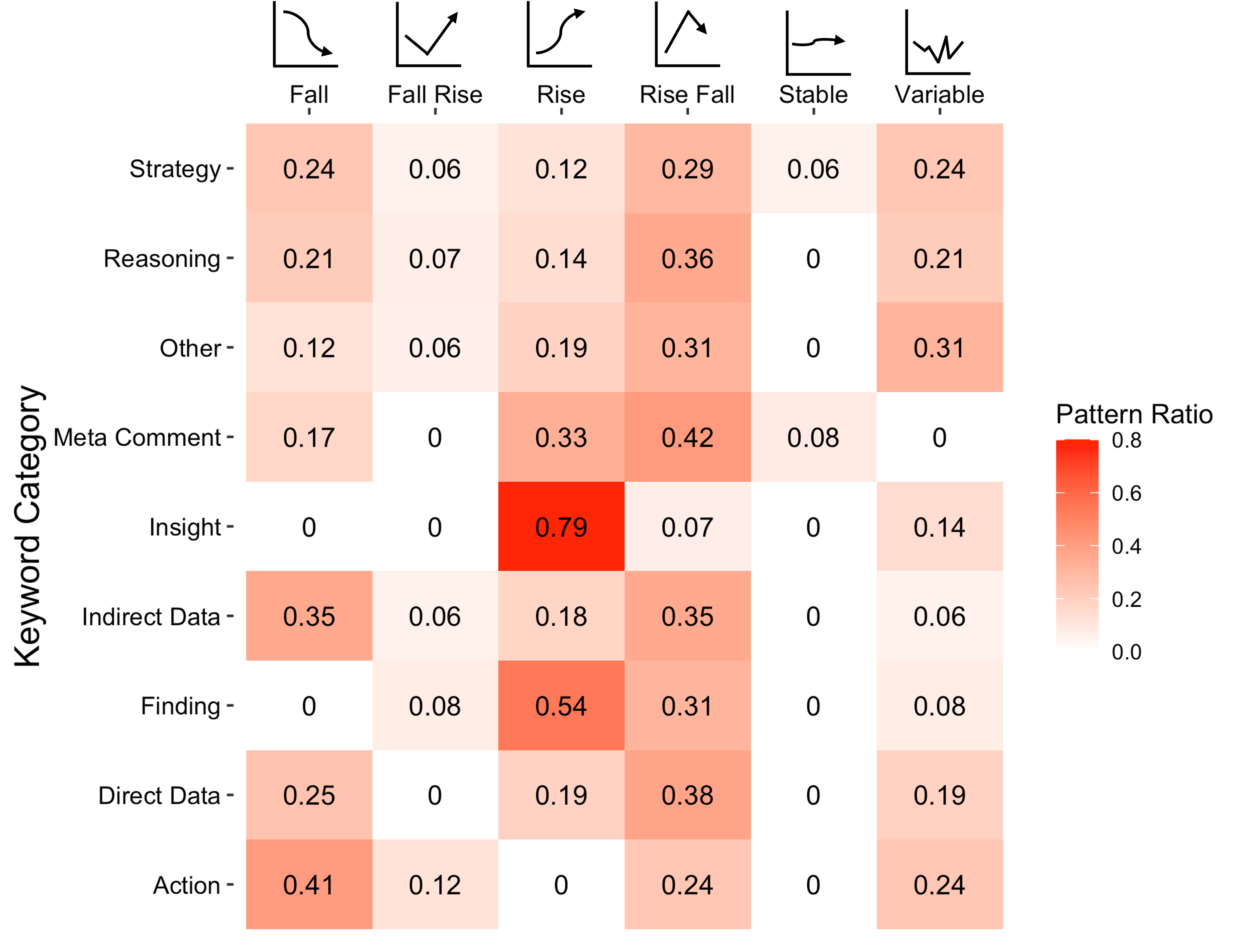}
    \caption{Heatmap displaying the ratio of each type of pattern per keyword category across all participants.
    The darkest and lightest shades are mapped to 0 and 0.8, respectively.
    }
    \label{fig:heatmap}
    \Description{This figure shows a heatmap depicting values for pattern ratio across different keyword categories.
    The patterns are shown in the horizontal axis, including \textit{fall}, \textit{fall rise}, \textit{rise}, \textit{rise fall}, \textit{stable}, and \textit{variable}.
    The keyword categories are shown in the vertical axis, including \textit{strategy}, \textit{reasoning}, \textit{meta comment}, \textit{insight}, \textit{indirect data}, \textit{direct data}, \textit{finding}, \textit{action}, and \textit{other}.
    The ratio of the patterns ranges from 0 to 0.80.
    }
\end{figure}

Figure~\ref{fig:heatmap} shows the color-coded heatmap of the ratio of each pattern per keyword category across all participants.
We categorize patterns as either (i) \textit{fall} (when the ratio for the keyword category decreases over time), (ii) \textit{rise} (when the ratio increases over time), (iii) \textit{fall-rise} (U shape), (iv) \textit{rise-fall} (inverted U shape), (v) \textit{stable} (when the ratio is fairly consistent over time), and (vi) variable (when there is no clear pattern or the change does not fit the other types).

The darkest red cell in figure~\ref{fig:heatmap} indicates that the \textit{Insight} category has a rising pattern throughout the analysis workflow.
This rise pattern corresponds to the fact that the analysis process will result in more insights later throughout the analysis. 
Similarly, the \textit{Findings} category has a rising pattern as commonly more results will be revealed throughout the analysis workflow.
In contrast, the \textit{action} category has a falling pattern, indicating that people usually spend more time using the visual tool facilities or exploring the content at the beginning of data analysis.
Closer to the end of the analysis process, the rate of reading, searching, and other kinds of action decreases as people spend more time forming deeper insights.

As seen in the \textit{stable} column of figure~\ref{fig:heatmap}, the majority of the categories do not have consistent patterns over time.
This result suggests that, when recommending summarization strategies, it is essential to adjust information composition over time rather than relying on a static composition.

\subsection{Reasons for Stage Changes}

Adding to the knowledge of how to summarize analysis history into smaller stages, our research also considered human criteria for identifying change points to separate the analysis history into segments.
The results of the affinity analysis enabled the identification of different reasons for stage changes between consecutive summary cards.
We observed that each reason derives from either significant findings from the previous stage, information on the current card, or, in some cases, both. 
The study data included a total count of 218 individual instances of analysis of reasons for switching to a new card in summarization.
The analysis identified eleven main themes emerging from our affinity diagram process. 
Here, we describe the empirically derived themes with percentages of observed occurrence (ordered from highest to lowest).
Note that given percentages are not mutually exclusive since some instances fit into multiple categories.

 \textbf{Key Data Finding [T1]} (22.94\%):
    Having the highest presence among themes, the reasonings in this theme are related to situations where finding data of significant importance would cause a switch to another summary stage.
    These findings can include a simple clue (e.g., ``switching to learn more about the two jets found in previous stage''), a data connection of interest (e.g., ``finding three linked articles and a common suspect in previous stage and moving to look into those findings''), or identification of a data item that prompts following questions.
    We observed that the key finding was the most important reason for our participants to wrap up this summary stage and shift to the next one.
    Often following a new data finding, it was common to observe a new set of purposeful actions in the next analysis stage (e.g., a sequence of new searches for a specific keyword in the next step or a period of data-access operations to follow up on the key findings).

    \textbf{Action-Based Reasons [T2]} (20.18\%):
    This theme includes reasons related to a change due to executing an action or using specific tool functions, as being directly related to \textit{action} keywords (refer to Table \ref{tab:code-set}).
    We observed that the action-based reasons either originated following a result of a \textit{key data finding} [T1] in the previous stage or were the reason to change to the next summary independent of the last stage.
    Some examples of action-based reasons from our sample analysis cases include switching to organizing documents, reading results after a keyword search, or searching keywords after finding essential data.

    \textbf{Goal-Oriented Exploration [T3]} (16.97\%):
    Reasons for this theme include some strategy-based scenarios with the common approach of going through data with a specific purpose rather than a more general exploration.
    For example, switching to searching for a specific term or switching to finding particular links to other items are among the reasons for the change in this theme.
    
    \textbf{Insight-Based Reasons [T4]} (13.30\%):
    This set of reasons appears when the reason behind the change is the analyst developing an \textit{insight}, i.e., a more profound and higher-level understanding or synthesis of findings, which would lead to a significant or clever result (refer to Table \ref{tab:code-set} for related \textit{insight} keywords).
    In this theme, we observed participants would switch to a new summary stage that resembles an understanding of cause and effect based on the identified relationships in the previous analysis stages.
    In other words, there is a switch to synthesizing ideas and findings to reach a conclusion.
    Some examples of this theme are ``switching to the conclusion and final insights'' or ``forming higher-level relations''.

    \textbf{Change of Item Focus [T5]} (11.01\%):
    Another set of reasons is based on the analyst's concentration on a discrete data component, item, or entity in a summary stage.
    We observed that participants would consider switching to analyzing a specific part of data as a reason to begin a new summary stage.
    Examples of this theme are switching to focus on a specific document or switching to find documents or information with particular data attributes.
    This theme is related to T1 in accounting for changes in attention to data, but T5 does not involve a notable finding or outcome.
    
    \textbf{Change of Topic Focus [T6]} (5.05\%):
    This theme is similar to T5, except that the analyst's focus changed drastically concerning the topic of investigation.
    This set of reasons appeared when the strategy was to shift the focus of attention between the summary stages, such as switching to focus on an entirely different topic or shifting the broader investigation direction.
    In other words, reasons attributed to topic changes are based on higher-level data changes rather than specific items.
    
    \textbf{Re-Examination [T7]} (3.67\%):
    In this theme, participants would switch to a new summary note card due to a change in the analyst's strategy, specifically involving revisiting previously gathered information. 
    This reason would indicate a point in the data analysis workflow where the analyst would look at the findings from a different perspective or refresh their previous connections of thoughts before moving forward with the following data analysis steps.
    
    \textbf{Initiation [T8]} (1.84\%):
    This theme includes transition reasons at the beginning of the analysis process after the analyst takes the earliest steps of their analysis.
    Some examples of this theme are changes following the exploration of the nearest or the first data document or clustering related data together.
    
    \textbf{Dead End [T9]} (1.84\%):
    This theme includes changes to another summary stage that occurs when the analysis process reaches a point where meaningful progress stops.
    Therefore, reaching a dead-end in that specific thread of analysis would lead to another stage that includes a different strategy or action by the analyst.
    
    \textbf{Completed Analysis Process [T10]} (1.38\%):
    Another reason participants switch to a different summary card includes cases involving merely completing a previous action or process.
    Separate from theme T2 (based on actions or using tool functions), T10 is based on stage transitions that happen when the analyst finishes an action or a series of actions and then moves on to begin a new series of analysis processes as a result of gathered knowledge from the prior finished process. 
    
    \textbf{Filtering [T11]} (1.38\%):  
    Reasons for stage changes under this theme are based on the analyst's transition to filter, refine, or re-organize the data items or findings. 
    Examples of this theme include separating important items from a larger set or ruling out irrelevant results.

While the themes above are directly derived and distinguished through the qualitative analysis, the main themes can be aggregated for more practical use.
We, therefore, used inductive thematic analysis~\cite{braun2013successful} to further examine the themes and categorized them into four higher-level categories, which are summarized in Table~\ref{tab:category} along with the aggregated proportion of occurrence.
Among the higher-level reasons for stage transition, the most common are \textbf{data-oriented reasons} related to specific data content given attention in the analysis.
The next most common is the aggregated category of \textbf{approach-based reasons}, which accounts for analysts' strategies and higher-order behaviors.
\textbf{Action-based reasons} for change are based on specific actions or execution of tool functions.
Lastly, \textbf{insight-based reasons} involve higher-level findings and inferences.

\begin{table*}[t!]
    \caption{Higher-level categories of themes for reasons for the decision to change to a new summary stage. 
    }
    \label{tab:category}
    \centering
    \resizebox{\textwidth}{!}{
    \small
    \rowcolors{2}{}{lightgray}
    \begin{tabular}{m{0.2\textwidth}m{0.45\textwidth}m{0.1\textwidth}m{0.25\textwidth}} \hline & \\[-1.5ex]
        Reasoning Category &
        Description &
        Ratio & \\ [1ex] \hline & \\[-1.5ex]
        
        Data-Oriented Reasons
        & Derived from the content of data sets and related to \textit{direct} and \textit{indirect} \textit{data} keywords. [T1, T5, T6]
        & \chartthree{39.00}  \\[1ex] & \\[-1.5ex]
        
        Approach-Based Reasons
        & Derived from higher-level behaviors and strategies, and related to \textit{strategy} keywords. [T3, T7, T8, T9, T10, T11]
        & \chartthree{27.06} \\[1ex] & \\[-1.5ex]
        
        Action-Based Reasons
        & Derived from actions done by the visual tool properties and related to \textit{action} keywords. [T2]
        & \chartthree{20.18}\\[1ex] & \\[-1.5ex]
        
        Insight-Based Reasons
        & Derived from higher-level findings and interpretations and related to \textit{insight} keywords. [T4]
        & \chartthree{13.30}\\ [1ex] \hline
    \end{tabular}
    }
\end{table*}

\section{Discussion}

We conducted an empirical user study to investigate users' strategies and patterns in summarizing analytic workflows into shorter, more meaningful chunks.
The results of the study contribute an empirical basis for the design of automated methods to aid in summarizing and visualizing data analysis workflows for reporting, presentation, and collaborative communication.
In this section, we discuss implications drawn from the findings on segmenting and summarizing analysis workflows.
The findings contribute to the broader goal of developing more effective tools and techniques for the automatic summarization of analytic provenance.

\subsection{Implications for Summary Content}

The qualitative analysis informs us about the content and composition of different types of provenance information that people use when constructing analysis summaries.
While the presented keyword categories were derived through coding and analysis of the study data, we can interpret keywords as they relate to broader types of analysis information.
Most notably, the large amount of data details in the summaries (based on the \textit{direct data} and \textit{indirect data} categories) demonstrate the importance of summarizing the data explored or analyzed across the stages of analysis.
The results show summaries also heavily incorporated \textit{actions} and \textit{strategy} information to explain the behavior used to analyze the data content.
Other key information about \textit{findings}, \textit{reasoning}, and \textit{insights} were also clearly included, but these appeared in lesser overall proportion and were more often used to summarize later stages of the analysis period.

For the goal of automating summarization visualization, an important realization from the results is that most of the information types (i.e., \textit{direct data}, \textit{indirect data}, and \textit{actions}) account for information that could largely be captured computationally based on observable user interaction events with associated data through analysis software.
These top three categories made up approximately 66\% of all summary keywords.
This is promising for the potential of automation to save a great deal of human effort in generating workflow summaries with a large amount of desired information.
However, some types of information, such as higher-level human reasoning and data interpretations, would be much more difficult to infer automatically.
While current algorithmic approaches may struggle to capture high-level human reasoning, integrating LLMs could help infer higher-order insights by generating more context-aware summaries that align better with human interpretations of data.

Although purely computational approaches are unlikely to perfectly capture and summarize human reasoning, workflow annotation or interactive methods that allow human modification of partially automated summaries remain promising.
These methods address the need to represent more ``human-oriented'' information, such as thinking, rationale, and insights.

Our study provides additional findings on the importance of communicating different types of workflow information at different stages of time.
As seen in Figure~\ref{fig:heatmap}, the composition of information types is inconsistent over time, as participants' summaries prioritized different forms of information at different times.
Most salient is the increase in the inclusion of analysis \textit{findings} and \textit{insights} over time; logically, there would not be findings to summarize early, and there would be more important findings as the analysis progresses.
In contrast, the results demonstrate how summaries emphasized information about \textit{actions} earlier, and inclusion dropped over time.

As another example, while specific details in the summaries about data items or entities were the most common type of information overall (Figure~\ref{tab:coding-scheme}), the temporal patterns in Figure~\ref{fig:heatmap} show the data details either (a) fell over time, with summaries providing more data details early on but dropping later to focus more on findings or insights, or (b) spiked over time with a peak somewhere in the middle of the analysis.
These patterns also align with expectations of known patterns in the human sensemaking process, where people tend to focus on exploring data and finding information early on, and then later, the focus shifts to gaining higher-level insights from information and developing a more comprehensive mental model~\cite{pirolli2005sensemaking}.
The variability in the composition of information over time suggests it may not be ideal for automated summarization approaches to rely on rules for static compositions, and adopting even basic temporal adjustments would be expected to improve the overall quality of summaries to better align with how people prefer to infer and explain the analysis process.

\subsection{Implications for Workflow Segmentation}
The results also inform decisions for how to segment time into smaller chunks to aid human understanding.
From the time results of card duration, participants had a clear preference for summarizing shorter periods (i.e., less than 3 minutes; see Figure~\ref{fig:numCards-participant}), but variance was high, and it is important to allow for longer periods based on the analysis behavior or lack of change in activity in a longer time.

The criteria for temporal chunking is closely related to the results for participants' reasons for changing to a new summary stage, as summarized in Table~\ref{tab:category}.
Notable and promising results for automation are the results of change points criteria, where the majority of reasons for stage changes (including data-oriented, approach-based, and action-based) stem from workflow events that can likely be detected computationally.
For instance, changes in user attention to different data content (the most common reason for change) can be detected by tracking data items accessed over time.
Prior work explored the use of topic modeling to identify temporal changes in the data content being analyzed~\cite{linder2016results}.
Many action-based reasons for change can similarly be directly detected based on interaction patterns, which could utilize simple rules to identify notable actions of interest (e.g., submitting a new search, opening a new tab to a different data view, or pausing interaction after an extended period of frequent activity)~\cite{linder2016results}, or use other established techniques for signal processing and change-point detection relying on anomaly detection, clustering, or supervised learning~\cite{aminikhanghahi2017survey}.

Changes in approach can also be detected since such a change is often associated with observable changes in user behavior and interaction patterns. 
However, changes in higher-level cognitive strategies or more nuanced changes in approach would likely remain challenging for automated detection.
Identifying changes in user insights is less straightforward but may also be associated with interaction behaviors or using explicit inputs such as note-taking or marking action items as complete from a task list.

\subsection{Implications for Tool Design}

One of the key implications of our findings is their potential application in the design of visual analytics tools that automate provenance summarization. 
Specific tools such as StoryFacets~\cite{Park_2021} and InsideInsights~\cite{mathisen2019insideinsights} can directly benefit from the user-driven strategies identified in this study. 
For instance, StoryFacets could leverage our results to inform decisions about the optimal number of segments or story pieces to start with, as well as the type of information to populate during the automatic creation of visual narratives. 
Similarly, InsideInsights can use these findings to automate the generation of lower-level insights, thus reducing the manual effort required in constructing detailed analytic narratives.

Furthermore, tools like Tessera~\cite{yan2021tessera}, which focuses on segmenting data analysis workflows based on interaction logs and data properties, could apply our recommendations to refine parameter weights and improve the selection of information for provenance summaries.
By adopting a more human-centered approach to these parameter decisions, tools like Tessera can align their automated outputs more closely with user expectations, enhancing both usability and interpretability.

An exciting direction for future research lies in exploring the potential of incorporating AI-driven techniques, such as large language models (LLMs), into provenance summarization processes. 
Given their success in natural language understanding, LLMs can play a crucial role in enhancing the automation of human-meaningful summaries by analyzing patterns in interaction logs and generating context-aware summaries. 
Combining LLMs with human annotation techniques could bridge the gap between purely algorithmic approaches and the nuanced understanding of user behavior, resulting in more effective and intuitive summarization tools.

However, while LLMs offer promise, their limitations must also be considered. 
Challenges such as domain-specific jargon, potential biases in summary generation, and the need for interpretability require further investigation to ensure that these models can deliver high-quality results in visual analytics contexts.

\subsection{Towards Generalization of Guidelines for Visual Workflow Summarization}

In aiming to support an easier understanding of complex analysis workflows, the rationale for segmenting experiences into summarized chunks is motivated by psychological foundations of how people conceptually build separable episodes or sub-goals in a chain of events to form a narrative~\cite{black1979episodes}.
In other words, narratives are an essential part of human expression, and people naturally tend to chunk a story into pieces and steps.
Our results regarding preferences and strategies for provenance summarization can provide an empirical starting point to support this human inclination towards narrative chunking, and we hypothesize that many of the fundamental strategies and considerations for narrative summarization may be independent of the topic of data analysis or the visual analytic system.

Still, an obvious challenge for automation is determining generalizable guidelines for visual summarization despite the fact that any empirical study must be grounded in specific cases.
Despite high-level commonalities in human strategies for storytelling, different data analysis scenarios may lend themselves to different summarization strategies.
To partially address this issue, our study was conducted with six separate provenance records with different analysts and spanning three separate datasets. 
However, the datasets were constrained to intelligence analysis scenarios with text documents to facilitate aggregation of results using a similar domain.
For extension of the research and further generalization of empirical foundations, it will be beneficial to conduct similar studies based on different datasets and analysis tools.
It will also be important to study different types of data analysts and end-users with varying levels of expertise (e.g., professional analysts) to contribute further knowledge of how experience influences summarization preferences.

Different people will naturally have different preferences for summarization and presentation of workflows, as the results of our study demonstrate.
For example, some participants summarized the analysis workflow into only a handful of stages (the lowest being 3), while the higher end included 19 stages.
Similarly, the level of detail or granularity for a stage varied greatly, as did individual compositions of different information types.
This aligns with findings from intelligence analysis expert interviews~\cite{block2024data}, where domain experts assessed a tool designed for summarizing and segmenting individual behaviors.
Notably, the experts did not express a clear preference for specific summarization formats.
A key takeaway from their reviews is the lack of consensus regarding the optimal length, level of detail, or format for process summaries.
Instead, they emphasized the importance of enabling users to customize or interactively explore summary details to match their unique contexts.

Expert preferences were highly dependent on contextual specifics and user needs, reinforcing the value of customizable summarization tools.
This suggests that while a general template may offer a useful starting point, customizable and interactive features are essential to accommodate the broad variability in user requirements and contexts.
The findings from our study, though based on particular cases, could inform initial customization efforts while allowing for later customization and flexibility to adapt to diverse user scenarios.

Consequently, we must accept that there is no such thing as a single ``correct'' summarization that will be perfect for everyone.
Summarization algorithms can be tuned to accommodate individual preferences and the needs of different organizations.
In addition, interactive modification and customization can provide further flexibility to refine summaries while still leveraging the time-saving benefits of automation.

\subsection{Limitations}

Our participant pool included students with varying levels of experience in data analysis and visualization. 
It's important to note that this group may not fully represent all users of analytic provenance tools. 
However, this sample remains valuable as it encompasses potential future users who may need to summarize analytic workflows, despite not being experts in these tools. 
Our research offers insights into how users with limited experience in provenance tools approach workflow summarization. 
In the future, it would be beneficial to expand this research to include domain experts to further test the generalizability of our findings.
Additionally, the study's focus on text analysis data may limit the immediate applicability of the results to other data types and analysis workflows. 
Future research should aim to replicate this study across various domains and data modalities to explore commonalities and differences in how users segment and summarize analytic workflows.

\section{Conclusion}

We investigated how users summarize and communicate analytic workflows in visual data analysis.
We conducted an empirical user study where the participants reviewed samples of provenance data using a visual analytics system and were asked to summarize the data analysis workflow into multiple stages.
By examining participants' natural behaviors and preferences when interacting with provenance data, we identified key parameters and strategies used to summarize data analysis processes.

Our findings provide valuable insights into human-centered approaches to provenance summarization, forming a solid empirical basis for developing automated techniques.
These results highlight the potential to improve visual analysis tools by incorporating user-driven strategies into the design of summarization algorithms, ultimately enhancing their capabilities in presentation and storytelling.

While the findings of our study were based on text analysis samples as a foundational starting point to inform summarization techniques, future research can expand on these findings to study commonalities and differences among different analysis contexts.
By building on this work, researchers can continue to explore new ways to bridge the gap between human-centered design and automated summarization in visual analytics.

\bibliographystyle{ACM-Reference-Format}
\bibliography{main}






\end{document}